\documentclass[runningheads]{llncs}
\usepackage{graphicx}
\usepackage{times}
\usepackage{latexsym}
\usepackage{amsmath}
\usepackage{amssymb}
\usepackage{listings}
\usepackage{booktabs}
\usepackage{multirow}
\usepackage{enumitem}

\usepackage{xcolor}

\begin{document}
\title{Cross-domain Retrieval in the Legal and Patent Domains: a Reproducibility Study}
%
%
\author{Sophia Althammer\and
Sebastian Hofst{\"a}tter \and
Allan Hanbury }
\authorrunning{S. Althammer et al.}
%
\institute{TU Wien, Austria\\
\email{\{first.last\}@tuwien.ac.at}}
\maketitle              
\begin{abstract}
Domain specific search has always been a challenging information retrieval task due to several challenges such as the domain specific language, the unique task setting, as well as the lack of accessible queries and corresponding relevance judgements.
In the last years, pretrained language models -- such as BERT -- revolutionized web and news search. Naturally, the community aims to adapt these advancements to cross-domain transfer of retrieval models for domain specific search.
In the context of legal document retrieval, Shao et al. propose the BERT-PLI framework by modeling the \textbf{P}aragraph-\textbf{L}evel \textbf{I}nteractions with the language model BERT.
In this paper we reproduce the original experiments, we clarify pre-processing steps and add missing scripts for framework steps, however we are not able to reproduce the evaluation results.
Contrary to the original paper, we demonstrate that the domain specific paragraph-level modelling does not appear to help the performance of the BERT-PLI model compared to paragraph-level modelling with the original BERT.
In addition to our legal search reproducibility study, we investigate BERT-PLI for document retrieval in the patent domain. We find that the BERT-PLI model does not yet achieve performance improvements for patent document retrieval compared to the BM25 baseline.
Furthermore, we evaluate the BERT-PLI model for cross-domain retrieval between the legal and patent domain on individual components, both on a paragraph and document-level.
We find that the transfer of the BERT-PLI model on the paragraph-level leads to comparable results between both domains as well as first promising results for the cross-domain transfer on the document-level.
For reproducibility and transparency as well as to benefit the community we make our source code and the trained models publicly available.

\end{abstract}
%


%
%
\section{Introduction}

Bringing the substantial effectiveness gains from contextualized language retrieval models from web and news search to other domains is paramount to the equitable use of machine learning models in Information Retrieval (IR).
The promise of these pre-trained models is a cross-domain transfer with limited in-domain training data.
Thus we investigate in this paper the document retrieval on two specific language domains, the legal and the patent domain, and study the transferability of the retrieval models between both domains.

In case law systems the precedent cases are a key source for lawyers, therefore it is essential for the lawyers' work to retrieve prior cases which support the query case.
Similarly in the patent domain, patent examiners review patent applications and search for prior art, in order to determine what contribution the invention makes over the prior art.
The recent advances in language modelling have shown that contextualized language models enhance the performance of information retrieval models in the web and news domain compared to traditional ad-hoc retrieval models \cite{Hofstaetter2019_osirrc,Hofstaetter2020_ecai}.
However for legal and patent retrieval we have a different task setting as the documents contain longer text with a mean of 11,100 words per document. In document retrieval every passage may be relevant, therefore in a high-recall setting such as ours it is crucial for the retrieval model to take the whole document into account. This is a challenge for contextualized language retrieval models, which are only capable of computing short passages with a length up to $512$ tokens \cite{earl,bertserini,dcbert}.

Recently, Shao et al.~\cite{bertpli} aimed to bring the gains of language modelling to legal document retrieval and tackle the challenge of long documents by proposing BERT-PLI, a multi-stage framework which models \textbf{P}aragraph-\textbf{L}evel \textbf{I}nteractions of queries and candidates with multiple paragraphs using BERT \cite{bert}.
The document-level relevance of each query and candidate pair is predicted based on paragraph-level interaction of the query and candidate paragraphs which are aggregated with a recurrent neural network (LSTM or GRU). The BERT-PLI model is trained in two stages: first, BERT is trained on a paragraph entailment task, and second the recurrent aggregation component is trained on a binary classification task.

In this paper we reproduce the results for the legal retrieval task. We found shortcomings in the description of the data pre-processing and evaluation methods, after a discussion with the authors of the original paper we could clarify how the evaluation results are achieved. As the published code is missing crucial parts, we re-implement the pre-processing, the first stage BERT fine-tuning as well as the retrieval with BM25 in the second stage and the overall evaluation.
Furthermore we analyze the ablation study of the original paper and answer the following research question:

\begin{itemize}[leftmargin=0.9cm]
    \item[\textbf{RQ1}] Does fine-tuning BERT on domain specific paragraphs improve the retrieval performance for document retrieval?
\end{itemize}

The original paper finds a $7-9\%$ performance improvement of the BERT-PLI model for legal retrieval, when fine-tuning BERT on the legal paragraphs.
Contrary to the original paper, we find that the paragraph-level modelling with BERT, fine-tuned on the domain specific paragraph-level modelling, does not appear to help the BERT-PLI model's performance on legal document retrieval. In line with that, we also demonstrate that the patent specific paragraph-level modelling harms the performance of the BERT-PLI model also for the patent retrieval task and remains a promising opportunity.

%
In order to analyze the proposed BERT-PLI model for another document retrieval task with long documents, we investigate following research question:

\begin{itemize}[leftmargin=0.9cm]
    \item[\textbf{RQ2}] To what extent is a BERT-PLI model, which is trained on patent retrieval, beneficial for document retrieval in the patent domain?
\end{itemize}

We find that the patent domain BERT-PLI model is outperformed by the BM25 baseline for the patent retrieval task. This shows that the document retrieval with BERT is not yet beneficial for the patent retrieval and stays a promising opportunity.

As the legal and patent documents come from similar language domains, it becomes an interesting question to what extent we can transfer the domain specific retrieval models from one to the other domain.
Especially because of the restricted accessbility of domain specific, labelled retrieval data there is the need for studying cross-domain transfer of document retrieval models.

\begin{itemize}[leftmargin=0.9cm]
    \item[\textbf{RQ3}] To what extent is cross-domain transfer on paragraph- and document-level of the domain specific BERT-PLI model between legal and patent domain possible?
\end{itemize}

We show that the transfer of the domain specific paragraph-level interaction modelling is possible between the legal and patent domain with similar performance of the retrieval model. Furthermore we find on the document-level transfer that the zero-shot application of a patent domain specific BERT-PLI model for the legal retrieval task achieves a lower performance than the BM25 baseline. Showing first promising results, the cross-domain transfer of retrieval models stays an open and exciting research direction.
Our main contributions are:
\begin{itemize}
    \item[$\bullet$] We reproduce the experiments of Shao et al. \cite{bertpli} and investigate shortcomings in the data pre-processing and model methods. Contrary to the paper we find that domain specific paragraph-level modelling does not appear to help the performance of the BERT-PLI model for legal document retrieval
    \item[$\bullet$] We train a domain specific BERT-PLI model for the patent domain and demonstrate that it does not yet outperform the BM25 baseline
    \item[$\bullet$] We analyze the cross-domain transfer of the BERT-PLI model between the legal and patent domain with first promising results
    \item[$\bullet$] In order to make our results available for reproduction and to benefit the community, we publish the source code and trained models at:\newline https://github.com/sophiaalthammer/bert-pli
\end{itemize}






\section{Methods}

\subsection{Task description}

Document retrieval in the legal and patent domain are specialized IR tasks with the particularity that query and candidates are long documents which use domain specific language.\newline
In legal document retrieval, the relevant documents are defined as the previous cases which should be noticed for solving the query case~\cite{colieesummary}, in other words which support or contradict the query document~\cite{bertpli}. The legal documents consist of long text containing the factual description of a case.\newline
Relevance in the patent domain is defined for the prior art search task \cite{clefipsummary}, i.e. it is the task to find documents in the corpus that are related to the new invention or describe the same invention. The patent documents consist of a title, an abstract, claims and a description as well as metadata like the authors or topical classifications. As we investigate retrieval and classification based on the textual information, we will only consider the textual data of the patent documents.

\subsection{BERT-PLI architecture overview}
\begin{figure}
    \centering
    \includegraphics[width=1.0\textwidth]{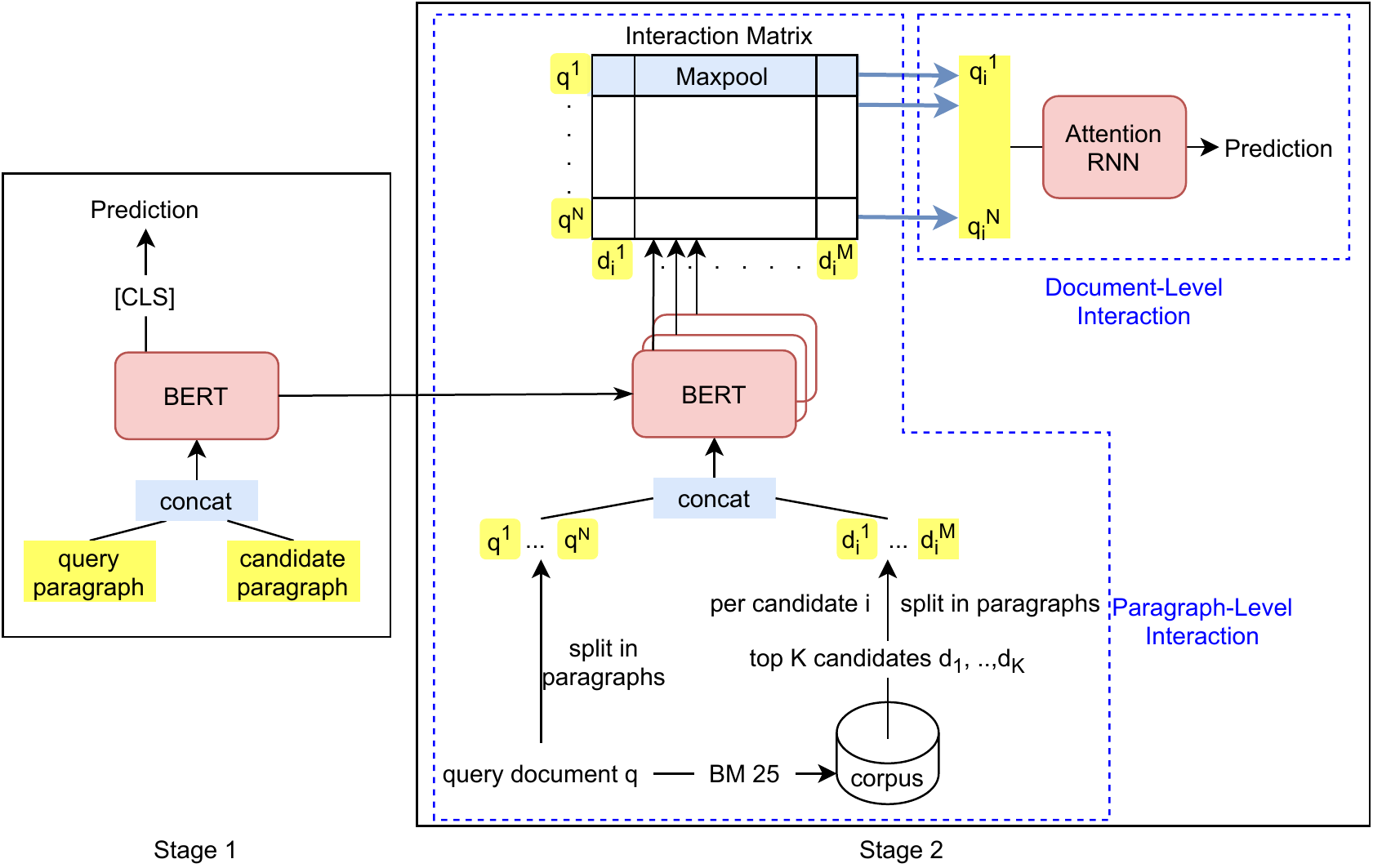}
        \caption{BERT-PLI Multistage architecture}
        \label{fig:arch}
\end{figure}
As the BERT model advanced the state-of-the-art in natural language processing and information retrieval, but has the restriction that it can only model the relation between short paragraphs, Shao et al.~\cite{bertpli} propose a multi-stage framework model using BERT for the retrieval of long documents which is illustrated in Figure \ref{fig:arch}. The training is separated into two stages. In stage 1, BERT is fine-tuned on a relevance prediction task on a paragraph-level. BERT takes the concatenated query and document paragraph as input and is then fine-tuned on predicting the relevance of the candidate paragraph to the query paragraph given the output vector of the special [CLS] token of BERT. Therefore this output vector is trained to be a relevance representation on a paragraph-level of the two concatenated input paragraphs.\newline
This fine-tuned BERT model is used in stage 2, where the full document retrieval with paragraph-level interaction modelling takes place. For a query document $q$ the top $K$ candidates are retrieved from a corpus using BM25~\cite{bm25}, and the query document as well as the top $K$ candidates are split into paragraphs. Then for each candidate $i \in {1, .., K}$ the first $N$ paragraphs of the query document and the first $M$ paragraphs of the candidate are concatenated and their relevance representation is calculated with the BERT model from stage 1. This yields an interaction matrix between the query and candidate paragraphs. An additional Maxpooling layer captures the strongest matching signals per query paragraph and yields a document-level relevance representation of the query and the candidate.
This document-level relevance representation is used to train an RNN model with a succeeding attention and fully-connected forward layer which we will refer to as Attention RNN. This Attention RNN yields the binary prediction of the relevance for the query and candidate document.

\subsection{Cross-domain evaluation approach}

\begin{figure}
    \centering
    \includegraphics[width=1.0\textwidth]{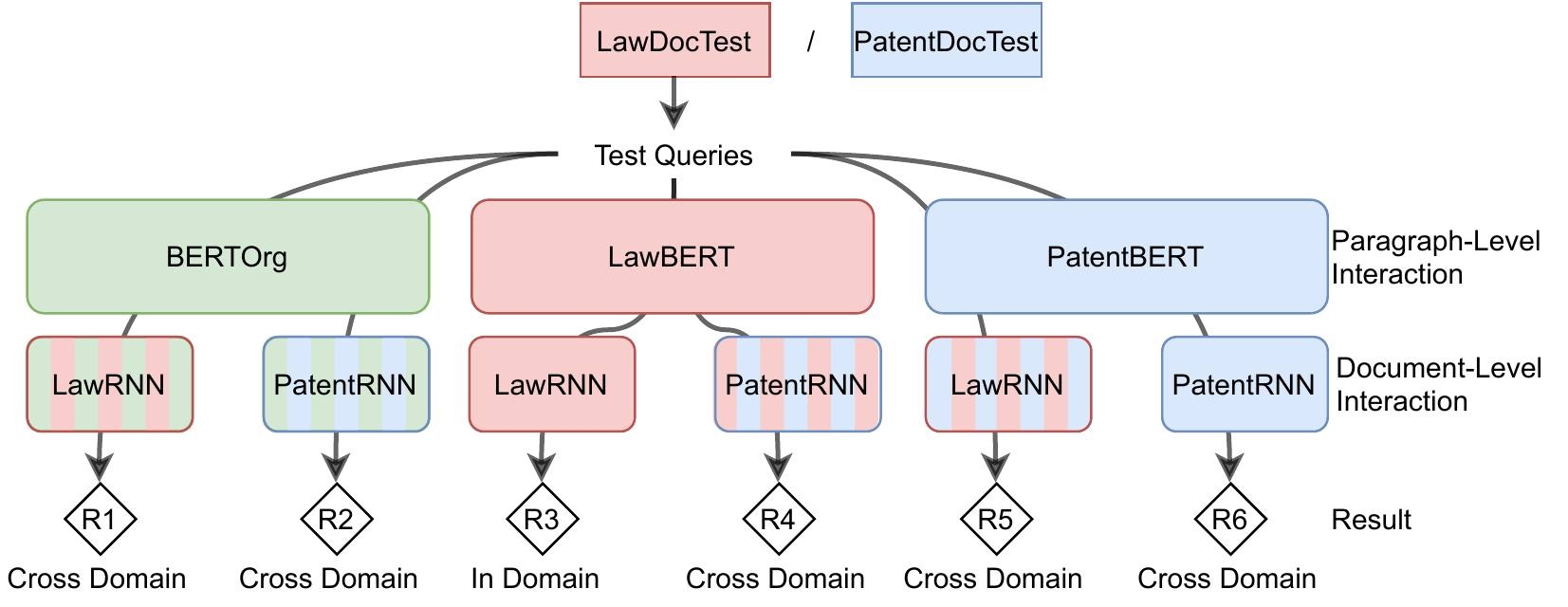}
        \caption{Cross-domain evaluation approach}
        \label{fig:crossdomain}
\end{figure}
In the first stage of the BERT-PLI framework the BERT model learns to model the paragraph-level interaction. For the two different domains we fine-tune the BERT model on a paragraph-level relevance prediction task, which yields the paragraph-level interaction \textbf{LawBERT} model for the legal and the \textbf{PatentBERT} model for the patent domain. In order to analyze the influence of the domain specific paragraph-level modelling, we compare the document retrieval models trained with the paragraph-level modelling of LawBERT or PatentBERT to document retrieval models trained on the paragraph-level modelling of the original BERT model. The paragraph-level modelling with the original BERT model is denoted with $\textbf{BERT}_\textbf{ORG}$ as in Figure \ref{fig:crossdomain}.
Based on these paragraph-level interaction representations we train an AttentionRNN on the legal as well as on the patent document-level retrieval task, which we denote with \textbf{LawRNN} or \textbf{PatentRNN} respectively. In order to isolate the impact of the different modelling of the paragraph-level interactions from LawBERT and PatentBERT, we additionally train an AttentionRNN on the patent document retrieval task given their LawBERT relevance representations and vice versa.\newline
We evaluate the resulting models on the legal or the patent test document retrieval set, namely \textbf{LawDocTest} or \textbf{PatentDocTest}. This process is visualized in Figure \ref{fig:crossdomain} and yields six evaluation results R1-6 for each test set. For example for LawDocTest, R3 is the in-domain evaluation result, whereas the other results denote cross domain evaluations.
For LawDocTest the results R1, R3 and R5 are all from LawRNN document retrieval, but the LawRNNs differ in the paragraph-level relevance representation they are trained with. Therefore comparison of the results R1, R3 and R5 on LawDocTest shows the transferability of the paragraph-level modelling between the legal and patent domains and the difference of domain-specific paragraph-level modelling to the non-domain specific modelling.
Furthermore to analyze the cross-domain transfer on the document-level, we compare the evaluation results of LawDocTest and PatentDocTest of R1 and R2, R3 and R4 as well as R5 and R6. This comparison shows the cross-domain transferability on the document-level as the LawRNN and PatentRNN share the same paragraph-level relevance representations, which they are trained on.

\section{Experiments}


\subsection{Datasets}


\subsubsection{Legal retrieval dataset}
Like Shao et al.~\cite{bertpli}, we use the legal retrieval collections from the COLIEE evaluation campaign 2019~\cite{colieesummary}, which consist of a paragraph-level and a document-level retrieval task. Both retrieval collections are based on cases from the Canadian case law system and are written in English. The paragraph-level task (COLIEE 2019 Task 2) involves the identification of a paragraph which entails the given query paragraph~\cite{colieesummary}. For this task the COLIEE evaluation campaign provides training and test queries with relevance judgements which we will refer to as \textbf{LawParaTrain} and \textbf{LawParaTest}. In the document-retrieval task (COLIEE 2019 Task 1) it is asked to find supporting cases from a provided set of candidate documents, which support the decision of the query document. As in the original paper we take $20\%$ of the queries of the training set as validation set, denoted with \textbf{LawDocVal}. We will refer to the training and test datasets for the document retrieval as \textbf{LawDocTrain} and \textbf{LawDocTest}. 

\subsubsection{Patent retrieval dataset}
For the patent retrieval queries and relevance judgments we use the datasets from the CLEF-IP evaluation campaign ~\cite{clefipsummary13} as they provide a patent corpus and training and test collections for patent retrieval tasks on the paragraph- and document-level. The tasks contain English, French and German queries, we only consider the English queries and candidates.
For the paragraph-level training and test collection we choose the provided queries and relevance judgements from the passage retrieval task starting from claims of the CLEF-IP 2013~\cite{clefipsummary13} where the participants are asked to find passages from patent documents which are relevant to a given set of claims. We refer to these datasets as \textbf{PatentParaTrain} and \textbf{PatentParaTest}.
As the document-level training and test collection we choose the queries and relevance judgements from the prior art candidate search from the CLEF-IP evaluation campaign 2011~\cite{clefipsummary} and refer to them as \textbf{PatentDocTrain} and \textbf{PatentDocTest}. As in the original paper, we take $20\%$ of the training set as validation set, denoted with \textbf{PatentDocVal}.
Both patent retrieval tasks retrieve paragraphs and documents from the patent corpus which consists of 3.5 million patent documents filed at the European Patent Office (EPO) or at the World Intellectual Property Office (WIPO).\newline
The dataset statistics can be found in Table \ref{table:datasetstats}.

\begin{table*}[]
\small
\centering
\caption{Statistics of the training and test set for the paragraph the document-level retrieval task}
\begin{tabular}{@{}lcccccccc@{}}
\toprule
& \multicolumn{2}{c}{LawPara} & \multicolumn{2}{c}{LawDoc} & \multicolumn{2}{c}{PatentPara} & \multicolumn{2}{c}{PatentDoc} \\
\cmidrule(lr){2-3} \cmidrule(lr){4-5} \cmidrule(lr){6-7} \cmidrule(lr){8-9}
           & Train & Test & Train & Test & Train & Test& Train & Test\\
\midrule
\# of queries & 181  & 41 & 285 & 61 & 44  & 42  & 351 & 100 \\
avg \# of candidates  & 32.12 & 32.19 & 200 & 200  & 3.5M  & 3.5M & 3.5M & 3.5M \\
avg \# relevant candidates & 1.12  & 1.02 & 5.21 & 5.41 & 43.52  & 76.3  & 3.27 & 2.85 \\
\bottomrule
\end{tabular}
\label{table:datasetstats}
\end{table*}

\subsection{Experiment setting}

\subsubsection{Stage 1: BERT fine-tuning}

In the first stage we fine-tune the BERT model\footnote{checkpoint from https://github.com/google-research/bert} on the paragraph-level relevance classification for either the legal domain or the patent domain to attain LawBERT and PatentBERT. As there was no code open-sourced for fine-tuning BERT, we use the HuggingFace transformers library\footnote{https://github.com/huggingface/transformers} and add the BERT fine-tuning script to the published code.\newline
For LawBERT we use the LawParaTrain as training and LawParaTest as test queries and relevance judgements. In order to use the queries and relevance judgements for a binary classification task, we consider the paragraph pairs of the query and one relevant candidate as positive samples. 
It was not stated clearly in the original paper how the paragraph pairs of negative samples are constructed, therefore we investigate this data pre-processing decision. We find that taking all paragraph pairs constructed of the query and a non-relevant paragraph from the paragraph candidates as negatives, yields comparable results for fine-tuning the BERT model on the legal domain as in the original paper. This negative sampling approach results in $3\%$ positive and $97\%$ negative samples in the training set.
The queries and paragraph candidates have less than $100$ words on average and are truncated symmetrically if they exceed the maximum input length of $512$ tokens of BERT.
For the training batch size we do a grid search and find that the F1-score of LawParaTest is the highest with a batch size of $2$ ($65.1\%$ F1-Score) instead of $1$ ($63.4\%$ F1-Score) after fine-tuning BERT for $3$ epochs on LawParaTrain, contrary to the original paper: they report the highest F1-score of $65.2\%$ without reporting the batch size. As stated in the original code, we assumed they used the batch size of $1$, due to our comparison we use a batch size of $2$ instead of $1$. After a remark of the original authors it turns out the original implementation was done with a batch size of $16$. For the learning rate we also do a grid search and find that the learning rate of $1e-5$ is optimal as in the original paper. As in the original paper, we fine-tune for $3$ epochs and we do the final fine-tuning of the LawBERT model on the merged training and test set. This is permissible as we train and evaluate the BERT-PLI model on LawDocTrain and LawDocTest, the LawParaTrain and LawParaTest sets are only used for fine-tuning LawBERT.\newline
For the PatentBERT fine-tuning we use the PatentParaTrain as training and PatentParaTest as test set.
We construct the negative paragraph pairs by sampling randomly paragraphs (which are not the relevant paragraph) from the documents which contain a relevant paragraph to a query paragraph. Here we sample randomly $5$ times the number of positive paragraphs as negatives, as otherwise the share of positive pairs is below $1\%$ and in order to have a similar ratio as for the legal domain. We do a grid search for the training batch size and learning rate and find that a batch size of $2$ with a learning rate of $2e-5$ yields the highest F1-score of $19.0\%$. We fine-tune PatentBERT solely on PatentParaTrain as it is common practice to hold out the test set.
\subsubsection{Stage 2: Document retrieval}

In stage 2 the first step is to retrieve relevant documents from the given set of candidates (in the legal domain) or from the whole corpus (in the patent domain). As it was not clearly stated in the original paper nor was there code published, how to employ the BM25 algorithm \cite{bm25} for this first step, we re-implement this step and use the BM25 algorithm \cite{bm25} with $k_1=0.9$ and $b=0.4$ implemented in the Pyserini toolkit\footnote{https://github.com/castorini/pyserini}. 
Furthermore we do a grid search for the input length to the BM25 algorithm and find that the top $K=50$ retrieval with input length of $250$ leads to similar recall scores as the original paper for the LawDocTrain set ($93.22\%$) and the LawDocTest ($92.23\%$). Here we only consider recall scores as in the original paper, as the focus of the first step BM25 retrieval is to retrieve all relevant cases for re-ranking for the training and test set.\newline
For patent document retrieval, the task is to retrieve relevant documents from the patent corpus with $3.5$ million documents. As in the patent document retrieval task only $3.27$ relevant patent documents per query document are contained and as the recall does not significantly increase when taking $K=50$ candidates, we choose the top $K=20$ from the BM25 retrieval, in order to have a similar ratio of positive and negative pairs as in the legal document retrieval for training the AttentionRNN. Here we find that the BM25 algorithm with the document input length of $250$ reaches the highest recall score of $9.42\%$ on PatentDocTrain compared to other document input lengths. Due to the low recall score of the retrieved documents on PatentDocTrain we add the relevant documents from the relevance judgements to PatentDocTrain and sample randomly non-relevant documents from the BM25 candidates for the training dataset, so that we have in total $20$ candidates. For PatentDocTest we retrieve the top $50$ candidates as in the original implementation where we reach a recall of $10.66\%$, but we do not add the relevant candidates after the BM25 retrieval step.
In order to reproduce the experiments for modelling the paragraph-level interaction and training the Attention RNN, we use the open-sourced repository\footnote{https://github.com/ThuYShao/BERT-PLI-IJCAI2020} of the original paper. As in the original paper we set the number of paragraphs of the query $N=54$ and the number of paragraphs of the candidate $M=40$ for legal and patent retrieval. The query and candidate documents are split up in paragraphs of $256$ tokens.
We model the paragraph-level interactions of LawDocTrain, LawDocTest, PatentDocTrain and PatentDocTest using LawBERT or PatentBERT or $\text{BERT}_{\text{ORG}}$. With these paragraph-level representations of each query and its candidate document we train an AttentionRNN network with either an LSTM~\cite{lstm} or a GRU network~\cite{gru} as RNN on classifying the relevance between the query and candidate document. The AttentionRNN trained on the LawDocTrain is denoted with LawRNN, on PatentDocTrain it is denoted with PatentRNN. For training the AttentionRNN we use the same hyperparameter as in the original implementation, except for the PatentBERT LawRNN configuration, where we find that the learning rate of $1e-4$ is better suited, when evaluated on the LawDocVal set.

\section{Evaluation and Analysis}

\subsection{In-domain evaluation for legal document retrieval (RQ1)}


Shao et al.~\cite{bertpli} evaluate their models using the binary classification metrics precision, recall and F1-Score on the whole test set. Furthermore they compare their model performance to the two best runs from the COLIEE 2019 denoted by the team names JNLP~\cite{jnlp} and ILPS~\cite{ilps}. As it was not clearly stated in the original paper, we assume that Shao et al. \cite{bertpli} evaluate the BERT-PLI models on the whole LawDocTest set with all $200$ given candidates per query. With the first retrieval step, the top $50$ query candidate pairs are retrieved for binary classification, therefore we assume the lower $150$ candidates classified as irrelevant. As in \cite{bertpli}, we use a cutoff value of $5$ for the evaluation of ranking algorithms like BM25, this means the top $5$ retrieved documents are classified as relevant, whereas the remaining $195$ are considered irrelevant.\newline
As Shao et al.~\cite{bertpli} evaluate in their published code the top $50$ candidates, we investigate the overall evaluation of our reproduced BERT-PLI models for all $200$ candidates with the precision, recall and F1-score using the SciKitlearn classification report \footnote{https://scikit-learn.org/stable/modules/generated/sklearn.metrics.classification\_report.html}. The results can be found in Table \ref{table:binclassificationeval}, we test the statistical significance compared to the BM25 baseline with the Student's paired, independent t-test \cite{ttestisbest2,ttestisbest}.
Comparing the evaluation results stated in the original paper and our evaluation results, we find that our reproduced BERT-PLI LawBERT LSTM and GRU model reach similar values.
On the effect of domain specific paragraph-level modelling on the legal case retrieval task (RQ1), the original paper reports a $7-9\%$ performance improvement for legal retrieval with the BERT-PLI model, when BERT is fine-tuned on the legal paragraph-level modelling compared to the original BERT.
Contrary to that, we find that the domain specific paragraph-level modelling does not appear to help the performance of the legal case retrieval. Our reproduced $\text{BERT}_\text{ORG}$ LawRNN GRU model outperforms all other BERT-PLI models except on the recall, however this shows that contrary to the findings in the original paper, the domain specific paragraph-level modelling does not always improve the performance of the BERT-PLI model.

\begin{table}[t]
\small
\centering
\caption{Precision, Recall and F1-Score comparison of Shao et al. \cite{bertpli} and our reproduction, BM25 cutoff value of 5 as in \cite{bertpli}, JNLP~\cite{jnlp} and ILPS~\cite{ilps} denote the best two runs of the COLIEE 2019, $^\dagger$ indicates statistically significant difference to BM25, $\alpha=0.05$}
    \begin{tabular}{@{}lccc@{}}
    \toprule
    Team/Model&Precision & Recall&F1-Score \\
    \midrule
    JNLP \cite{jnlp}& 0.6000& 0.5545&0.5764\\
    ILPS \cite{ilps}& 0.68&0.43&0.53\\
    $\text{BERT}_{\text{ORG}}$ LawRNN LSTM \cite{bertpli} &0.5278&0.4606&0.4919\\
    $\text{BERT}_{\text{ORG}}$ LawRNN GRU \cite{bertpli} &0.4958&0.5364&0.5153\\
    LawBERT LawRNN LSTM \cite{bertpli} &0.5931&0.5697&0.5812\\
    LawBERT LawRNN GRU \cite{bertpli}&\textbf{0.6026}&\textbf{0.5697}&\textbf{0.5857}\\
    \midrule
    Reproduction \\
    \midrule
    BM25 (cutoff at 5) & 0.5114 & 0.5360 & 0.5234\\
    Repr $\text{BERT}_{\text{ORG}}$ LawRNN LSTM &$0.7053^\dagger$&$0.5017^\dagger$&$0.5863^\dagger$\\
    Repr $\text{BERT}_{\text{ORG}}$ LawRNN GRU &$\textbf{0.8972}^\dagger$&$0.4501^\dagger$&$\textbf{0.5995}^\dagger$\\
    Repr LawBERT LawRNN LSTM &$0.8620^\dagger$&$0.4295^\dagger$&$0.5733^\dagger$\\
    Repr LawBERT LawRNN GRU &$0.3826^\dagger$&$\textbf{0.6838}^\dagger$&$0.4907^\dagger$\\
    \bottomrule
    \end{tabular}
    \label{table:binclassificationeval}
\end{table}

\subsection{In-domain evaluation for patent document retrieval (RQ2)}

In order to investigate the applicability of the BERT-PLI model for information retrieval in the patent domain, we evaluate the PatentBERT PatentRNN models trained on PatentDocTrain. The results can be found in Table \ref{table:evalpatent}, now we analyze the in-domain evaluation for the PatentBERT PatentRNN models on PatentDocTest.
This shows that the in-domain, patent BERT-PLI model is not beneficial for patent document retrieval, as it is outperformed by the BM25 baseline on all metrics. We reason this could be due to the number of considered query and candidate paragraphs ($N$ and $M$), which is fit to the legal retrieval but not to the patent retrieval and could be unsuitable for patent retrieval as PatentDocTrain and PatentDocTest contain on average more paragraphs than LawDocTrain and LawDocTest. This demonstrates that the document retrieval with contextualized language models for the patent domain is not yet beneficial and needs to be taken under further investigation. In line with the findings regarding RQ1 for the legal document retrieval, we find that the paragraph-level modelling with the PatentBERT model impairs the performance of the document retrieval compared to the paragraph-level modelling with $\text{BERT}_\text{ORG}$. This shows that the domain specific paragraph-level modelling is not always beneficial for BERT-PLI for the legal and patent document retrieval.

\begin{table}[t]
\small
\centering
\caption{In-domain and cross-domain evaluation on the legal and patent document retrieval test set, in-domain evaluation for LawBERT LawRNN models on LawDocTest and PatentBERT PatentRNN on PatentDocTest, R1-6 denote the result numbers from Figure \ref{fig:crossdomain}, $^\dagger$ indicates statistically significant difference to BM25, $\alpha=0.05$}
    \begin{tabular}{@{}llcccccc@{}}
    \toprule
     && \multicolumn{3}{c}{LawDocTest} & \multicolumn{3}{c}{PatentDocTest} \\
    \cmidrule(lr){3-5} \cmidrule(lr){6-8} 
    Model && Prec  & Rec & F1 & Prec  & Rec & F1  \\
    \midrule
    In-domain & &&& &\multicolumn{1}{|c}{}&&\\
    \midrule
    BM25 (cutoff at 5) && 0.5114 & 0.5360&0.5234&\multicolumn{1}{|c}{\textbf {0.0500}}&\textbf{0.3968}&\textbf{0.0888}\\
    \multirow{2}{*}{LawBERT LawRNN (R3)} & \multicolumn{1}{c}{LSTM} & \multicolumn{1}{c}{$\textbf{0.8620}^\dagger$} & \multicolumn{1}{c}{$0.4295^\dagger$} & \multicolumn{1}{c}{$\textbf{0.5733}^\dagger$} & \multicolumn{1}{|c}{$0.0207^\dagger$} & \multicolumn{1}{c}{$0.4761^\dagger$} & \multicolumn{1}{c}{$0.0398^\dagger$} \\
                                 & \multicolumn{1}{c}{GRU}  & \multicolumn{1}{c}{$0.3826^\dagger$} & \multicolumn{1}{c}{$\textbf{0.6838}^\dagger$} & \multicolumn{1}{c}{$0.4907^\dagger$} & \multicolumn{1}{|c}{$0.0181^\dagger$} & \multicolumn{1}{c}{$0.4444^\dagger$} & \multicolumn{1}{c}{$0.0349^\dagger$} \\
    \multirow{2}{*}{PatentBERT PatentRNN (R6)} & \multicolumn{1}{c}{LSTM}  & \multicolumn{1}{c}{$0.7500^\dagger$} & \multicolumn{1}{c}{$0.2268^\dagger$} & \multicolumn{1}{c}{$0.3482^\dagger$}   & \multicolumn{1}{|c}{$0.0365^\dagger$} & \multicolumn{1}{c}{$0.1904^\dagger$} & \multicolumn{1}{c}{$0.0613^\dagger$}   \\
                                 & \multicolumn{1}{c}{GRU}  & \multicolumn{1}{c}{$0.1153^\dagger$} & \multicolumn{1}{c}{$0.0412^\dagger$} & \multicolumn{1}{c}{$0.0607^\dagger$} & \multicolumn{1}{|c}{$0.0416^\dagger$} & \multicolumn{1}{c}{$0.1904^\dagger$} & \multicolumn{1}{c}{$0.0683^\dagger$} \\
    \midrule
    Cross-domain & &&& &\multicolumn{1}{|c}{}&&\\
    \midrule
    \multirow{2}{*}{LawBERT PatentRNN (R4)} & \multicolumn{1}{l}{LSTM}  & \multicolumn{1}{c}{$0.1103^\dagger$} & \multicolumn{1}{c}{$0.5292^\dagger$} & \multicolumn{1}{c}{$0.1826^\dagger$}  & \multicolumn{1}{|c}{$0.0277^\dagger$} & \multicolumn{1}{c}{$0.1587^\dagger$} & \multicolumn{1}{c}{$0.0472^\dagger$} \\
                                 & \multicolumn{1}{c}{GRU}  & \multicolumn{1}{c}{$0.0961^\dagger$} & \multicolumn{1}{c}{$0.2749^\dagger$} & \multicolumn{1}{c}{$0.1424^\dagger$} & \multicolumn{1}{|c}{$0.0246^\dagger$} & \multicolumn{1}{c}{$0.1904^\dagger$} & \multicolumn{1}{c}{$0.0436^\dagger$} \\
    \multirow{2}{*}{PatentBERT LawRNN (R5)} & \multicolumn{1}{c}{LSTM} & \multicolumn{1}{c}{$0.8000^\dagger$} & \multicolumn{1}{c}{$0.4673^\dagger$} & \multicolumn{1}{c}{$0.5900^\dagger$}  & \multicolumn{1}{|c}{$0.0188^\dagger$} & \multicolumn{1}{c}{$0.3650^\dagger$} & \multicolumn{1}{c}{$0.0357^\dagger$}   \\
    & \multicolumn{1}{c}{GRU}   & \multicolumn{1}{c}{$0.5460^\dagger$} & \multicolumn{1}{c}{$0.5704^\dagger$} & \multicolumn{1}{c}{$0.5579^\dagger$} & \multicolumn{1}{|c}{$0.0233^\dagger$} & \multicolumn{1}{c}{$0.5555^\dagger$} & \multicolumn{1}{c}{$0.0448^\dagger$}   \\
    \multirow{2}{*}{BERTOrg PatentRNN (R2)} & \multicolumn{1}{l}{LSTM}  & \multicolumn{1}{c}{$0.0000^\dagger$} & \multicolumn{1}{c}{$0.0000^\dagger$} & \multicolumn{1}{c}{$0.0000^\dagger$}   & \multicolumn{1}{|c}{$0.0602^\dagger$} & \multicolumn{1}{c}{$0.0793^\dagger$} & \multicolumn{1}{c}{$0.0684^\dagger$}  \\
                                 & \multicolumn{1}{c}{GRU}  & \multicolumn{1}{c}{$0.0000^\dagger$} & \multicolumn{1}{c}{$0.0000^\dagger$} & \multicolumn{1}{c}{$0.0000^\dagger$} & \multicolumn{1}{|c}{$\textbf{0.0769}^\dagger$} & \multicolumn{1}{c}{$0.0952^\dagger$} & \multicolumn{1}{c}{$0.0851^\dagger$}  \\
    \multirow{2}{*}{BERTOrg LawRNN (R1)} & \multicolumn{1}{c}{LSTM}  & \multicolumn{1}{c}{$0.7053^\dagger$} & \multicolumn{1}{c}{$0.5017^\dagger$} & \multicolumn{1}{c}{$0.5863^\dagger$} & \multicolumn{1}{|c}{$0.0160^\dagger$} & \multicolumn{1}{c}{$\textbf{0.8095}^\dagger$} & \multicolumn{1}{c}{$0.0314^\dagger$}  \\
                                 & \multicolumn{1}{c}{GRU} & \multicolumn{1}{c}{$\textbf{0.8972}^\dagger$} & \multicolumn{1}{c}{$0.4501^\dagger$} & \multicolumn{1}{c}{$\textbf{0.5995}^\dagger$}  & \multicolumn{1}{|c}{$0.0199^\dagger$} & \multicolumn{1}{c}{$0.4285^\dagger$} & \multicolumn{1}{c}{$0.0381^\dagger$}  \\
    \bottomrule
    \end{tabular}
    \label{table:evalpatent}
\end{table}

\subsection{Cross-domain evaluation (RQ3)}

In order to analyze the cross-domain retrieval between the legal and patent domain, we evaluate each model on LawDocTest and PatentDocTest set as illustrated in Figure \ref{fig:crossdomain} and compare for each test set the performance of the different models in order to gain insights about the transferability of the models between the legal and patent retrieval task and on the paragraph as well as on the document-level.\newline
Analyzing the cross-domain transfer on the paragraph-level for LawDocTest, we see in Table \ref{table:evalpatent} that the performance is similar for the LawRNNs when modelling the paragraph-level interaction with PatentBERT instead of LawBERT. An interesting result is the performance of the PatentBERT PatentRNN LSTM model, which was not trained on modelling legal paragraph-interactions nor legal document retrieval, but performs well on LawDocTest, however it does not outperform the domain independent BM25 baseline. On the document-level we see that the PatentRNN models have on average a $40\%$ lower F1-Score than the LawRNN models with the same paragraph-level modelling, although we see a positive effect of modelling the paragraph-level interactions with $\text{BERT}_{\text{ORG}}$ instead of LawBERT or PatentBERT.\newline
For the cross-domain evaluation on PatentDocTest, we find that each BERT-PLI model is outperformed by the BM25 baseline, except for the precision of the $\text{BERT}_{\text{ORG}}$ PatentRNN models and the recall of the $\text{BERT}_{\text{ORG}}$ LawRNN models.
On the document-level transfer we see a consistent performance improvement of the PatentRNN models compared to the LawRNN models independent of the paragraph-level modelling, which leads to the conclusion that the domain specific training for patent document retrieval is beneficial here. On a paragraph-level transfer we can see a similar performance of the LawRNN models, independent of the paragraph-level modelling. For the PatentRNN models we find that the paragraph-level modelling with $\text{BERT}_{\text{ORG}}$ outperforms the modelling with PatentBERT and LawBERT.

\section{Related Work}

There are numerous evaluation campaigns for patent \cite{clefipsummary} and legal retrieval \cite{fireaila,treclegal,colieesummary} with the goal to create and provide queries and relevance judgements for domain-specific retrieval and with this promote research in legal and patent IR.
For legal retrieval, Cormack et al. \cite{cormackgrossman1} evaluate continuous, simple active and passive learning models in the TREC legal evaluation campaign \cite{treclegal} and propose an autonomous active learning framework \cite{cormackgrossman2}.
In the COLIEE evaluation campaign, Rossi et al. \cite{ilps} combine text summarization and a generalized language model to predict pairwise relevance for the legal case retrieval task, whereas Tran et al. \cite{jnlp} apply a summarization method and the extraction of lexical features.
In the patent retrieval evaluation campaign CLEF-IP \cite{clefipsummary13}, Piroi et al. \cite{florinapatent} report different approaches using the probabilistic BM25 model \cite{bm25} as well as SVM-classifier trained on pretrained word-level representations.\newline
As the language model BERT \cite{bert} advanced the state-of-the-art in language modeling, there are numerous approaches to apply BERT to IR tasks \cite{lee-etal-2019-latent,bertserini} and for cross-domain IR for web and news search \cite{crossdomainevidence} as well as for biomedical search \cite{macavaney-etal-2020-sledge,xiong2020cmt}.

\section{Conclusion and Future Work}




We reproduced the BERT-PLI model of Shao et al. \cite{bertpli} for the legal document retrieval task of the COLIEE evaluation campaign 2019 \cite{colieesummary}. We have addressed shortcomings of the description of the data pre-processing and the second stage retrieval, which we investigated and for which we complemented the published code.
Contrary to the original paper, we find that modelling the paragraph-level interactions with a BERT model fine-tuned on the domain does not appear to help the performance of the BERT-PLI model for document retrieval compared to modelling the paragraph-level interactions with the original BERT model.
Furthermore we have analyzed the applicability of the BERT-PLI model for document retrieval in the patent domain, but we find that the BERT-PLI model does not yet improve the patent document retrieval compared to the BM25 baseline. We reason that the optimal number of query and candidate paragraphs to be considered for the interaction modelling could be a decisive hyperparameter to take into account. However bringing the gains from contextualized language model to patent document retrieval stays an open problem.
We have investigated to what extend the BERT-PLI model is transferable between the legal and patent domain on the paragraph and document-level by evaluating the cross-domain retrieval of the BERT-PLI model.
We show that the cross-domain transfer on the paragraph-level yields comparable performance between the legal and the patent domain. Furthermore the comparison on the document-level transfer shows first promising results when applying the BERT-PLI model trained on the patent domain to the legal domain.
How to bring the benefits of contextualized language models to domain-specific search and how to transfer retrieval models across different domains remain open and exciting questions.

\section*{Acknowledgements}
This work was supported by the EU Horizon 2020 ITN/ETN on Domain Specific Systems for Information Extraction and Retrieval (H2020-EU.1.3.1., ID: 860721).

%
%
%

\bibliographystyle{splncs04}
\bibliography{mybibliography}
%

\end{document}